\documentclass[12pt,preprint]{aastex}
\usepackage{natbib}
\usepackage{subfigure}
\usepackage{times}
\begin{document}

\title{Detection of Gamma-Ray Emission in the Region of the \\ Supernova Remnants G296.5+10.0 and G166.0+4.3}

\author{Miguel Araya}
\affil{Space Research Centre (CINESPA), Universidad de Costa Rica}
\affil{San Jos\'e 2060, Costa Rica}
\email{miguel.araya@ucr.ac.cr}

\begin{abstract}
52 months of accumulated observations by the Large Area Telescope onboard the \emph{Fermi Gamma-ray Space Telescope} in the region of the supernova remnants G296.5+10.0 (PKS 1209-51/52) and G166.0+4.3 (VRO 42.05.01) are analyzed. GeV emission is detected coincident with the position of the sources at the $\simeq$ 5$\sigma$ and 11$\sigma$ levels above the background, respectively, for the best-fit spectral and spatial scenarios. The gamma-ray spectrum of the sources can be described with a power-law in energy. G166.0+4.3 shows a soft GeV spectrum while that of G296.5+10.0 is flat (in the $\nu F_{\nu}$ representation). The origin of the gamma-ray emission from the sources is explored. Both leptonic and hadronic mechanisms can account for the high-energy emission from G296.5+10.0, while a leptonic scenario is preferred for G166.0+4.3.
\end{abstract}


\section{Introduction}
Supernova remnants (SNRs) are known to be high-energy sources exhibiting non-thermal photon spectra from radio to gamma-rays. High-energy photon emission may result from interactions of high-energy protons with ambient nuclei resulting in the production of neutral pions that decay to gamma-rays. This is generally referred to as the hadronic scenario. Leptonic mechanisms for the production of gamma-rays in SNRs include inverse Compton up-scattering (IC) of low energy photons by high-energy electrons and non-thermal bremsstrahlung emission \cite[e.g.,][]{gaisser1998} from high-energy electrons interacting with ambient particles.

Certainly, high-energy electrons are known to be accelerated in SNRs from their radio and X-ray synchrotron emission; the latter is usually associated to the forward shock of young remnants \citep[e.g.,][]{got01,ber02,hwa02,rho02,lon03,vink03,ber04}. Particles are thought to gain energy through first order Fermi acceleration \citep{bel78,bla87}, also known as diffusive shock acceleration (DSA) after crossing the shock front of the supernova explosion, which occurs many times as the particles move in the presence of magnetic fields that result either from shock compression of the interstellar field or from amplification by cosmic ray instabilities \citep[see][for a recent review]{schure2012}.

It is thought that a considerable ($1-10\%$) fraction of the SNR energy can be transferred to particles via DSA, which is one of the reasons why they are considered the main source of Galactic cosmic rays. However, identifying and separating the emission from leptonic and hadronic components is a challenging task. Some SNRs interacting with molecular clouds, such as W28, W49B, W51C and G8.7-0.1, show gamma-ray emission that seems to favour a hadronic origin \citep{abdo2010b,abdo2010c,abdo2009a,ajello2012}. Other studies show that some young SNRs with a relatively hard GeV photon spectrum are probably leptonic-dominated, such as RX J1713.7-3946 \citep{abdo2011} and RX J0852.0-4622 \citep{tanaka2011}. The softer spectrum of other young sources such as Cas A \citep{abdo2010a,araya2010} and Tycho SNR \citep{giordano2012} might also favour a hadronic origin. Recent observations of the SNRs IC 443 and W44 seem to have confirmed the presence of cosmic ray protons (more generally, ions) from their characteristic spectrum below $\sim 100$ MeV \citep{ackermann2013}.

Observations from the recently-launched \emph{Fermi} satellite \citep{atw09} have contributed to form a more consistent picture of gamma-ray emission from SNRs and particle acceleration in SNR shocks \cite[e.g.,][]{caprioli2012}. Despite the advances in the understanding of SNR properties, important limitations are still common for gamma-ray studies. For example, the broad point spread function (PSF) of the Large Area Telescope (LAT) onboard \emph{Fermi} and the high Galactic radiation background often present challenges for data analysis. In this paper, a study of \emph{Fermi} LAT observations is carried out for two remnants located outside the Galactic plane, where the expected Galactic diffuse level is lower.

The sources are G296.5+10.0 and G166.0+4.3, shell-type SNRs with radio extensions $90'\times65'$ and $55'\times35'$, respectively. The distance adopted here for G296.5+10.0 is 2.1 kpc \citep{giacani2000}. In the case of G166.0+4.3, a distance estimate of $4.5 \pm 1.5$ kpc \citep{landecker1989} is used.

G166.0+4.3 is composed of two parts as seen in radio images: a regular shell to the East and a region expanding into a low density medium (the extended `wing') to the West. The morphology of this remnant possibly results from the shock encountering a density discontinuity \citep{pineault1985} in the interstellar medium (ISM). The X-ray emission is only present in the interior and peaks towards the West wing \citep{burrows1994}, where the shock has encountered denser material \citep{landecker1989}. HI observations have shown ISM features interacting with the SNR \citep{landecker1989}.

G296.5+10.0 is a barrel-shaped SNR. The detected X-ray emission is in good correspondence with the radio. The morphology of barrel-shaped SNRs might be the result of their interactions with the ISM material or the magnetic field in the ISM, or result from the intrinsic properties of the outburst and later interaction with the ISM \cite[e.g.,][]{kesteven1987}. Located relatively far above the Galactic plane, G296.5+10.0 is possibly surrounded by low-density, uniform ISM. However, HI observations have revealed three clouds that are associated with the SNR \citep{giacani2000}: a long, broad structure of size $1^{\circ}\times 25'$ to the northeast, a cloud along the southwestern limb, near the Galactic coordinates $(l,b) \sim (296^{\circ}.00,+9^{\circ}.50)$ (volume density $\sim 13$ cm$^{-3}$), and the HI cloud across the eastern limb close to the brightest filaments near $(l,b) \sim (296^{\circ}.67,+9^{\circ}.67)$ (density $\sim 13$ cm$^{-3}$).

The structure of this paper is as follows. In Section \ref{observations} the gamma-ray data reduction is discussed, paying attention to the morphology and spectral properties of the emission detected in the direction of the SNRs. In Section \ref{model}, the non-thermal spectral energy distributions (SED) are modeled with different emission mechanisms. The discussion of results and final remarks are given in Section \ref{discussion}.

\section{Observations}\label{observations}
\subsection{Radio data}
The radio data points for G296.5+10.0 and G166.0+4.3 are obtained from the literature \citep{milne1994,leahy2005,leahy2006}. The radio spectrum of the sources used here can be accounted for by optically-thin synchrotron emission from cosmic ray electrons. The particle spectrum responsible for the radio emission is a power-law in energy ($\propto \epsilon^{-s}$ with $\epsilon$ the particle energy). From standard results concerning the synchrotron emission by a power-law population of electrons, the radio data imply $s \simeq 2$ and $\simeq 1.7$ for G296.5+10.0 and G166.0+4.3, respectively.

\subsection{\emph{Fermi} LAT data}
\emph{Fermi} LAT data taken between 04 August 2008 and 24 January 2013 were analyzed with the standard software \emph{ScienceTools} version v9r27p1\footnote{See http://fermi.gsfc.nasa.gov/ssc .} released April 18, 2012. Several selection criteria are applied to events, including the selection of events with high probability of being gamma-rays (the so-called Pass 7 \emph{Source} class) and with a reconstructed zenith angle less than 100$^{\circ}$ to avoid contamination from gamma rays from Earth's limb. Time intervals when the spacecraft is within the range of rocking angles used during nominal sky-survey observations (the rocking angle is less than 52$^{\circ}$) are also selected. The spectral analysis is further restricted above 200 MeV to avoid uncertainties in the effective area and broad PSF at lower energies, and below 100 GeV due to limited statistics. The same data selection criteria were used for regions near SNRs G296.5+10.0 and G166.0+4.3.

Events within a square region of 14$^{\circ}\times$14$^{\circ}$ of the catalogued positions of the SNRs G296.5+10.0 and G166.0+4.3, RA (J2000)= 12$^h$09$^m$40$^s$, Dec (J2000)= -52$^\circ$25$'$00$''$ and RA (J2000)= 05$^h$26$^m$30$^s$, Dec (J2000)= 42$^\circ$56$'$00$''$, respectively, are included in the analysis. This is necessary to account for the large PSF of the LAT. The emission model that is used in the analysis includes the positions and spectral parameter values of the sources within this region that are found in the LAT 2-year Source Catalog \citep{nolan2012}. In the case of the region containing G296.5+10.0 nearby extended sources (known as Cen A and MSH 15-52) are modeled with spatial templates provided within the \emph{ScienceTools}. The data are binned in sky coordinates with the tool \emph{gtbin} in square bins of size 0.$^\circ$1 to construct count maps for visualization.

The spectral and spatial properties of \emph{Fermi} LAT data are explored by means of a maximum likelihood analysis\footnote{See http://fermi.gsfc.nasa.gov/ssc/data/analysis/scitools/binned\_likelihood\_tutorial.html .} using the tool \emph{gtlike}. The likelihood is defined as the probability of obtaining the data given an input spatial and spectral model for the sources. The starting point for the fitting procedure is obtained from the LAT 2-year Source Catalog, as mentioned above. The currently released instrument response functions ($P7SOURCE\_V6$) are used throughout the analysis, as well as the latest galactic and extragalactic diffuse background components (as specified in the files \emph{gal\_2yearp7v6\_v0.fits} and \emph{iso\_p7v6source.txt}, respectively). The spectral parameters of the catalogued sources beyond 7$^\circ$ from the position of G296.5+10.0 and G166.0+4.3 are kept fixed to the values reported in the catalog. The fit is performed with the optimizer NEWMINUIT until convergence is achieved.

In the case of the SNR G166.0+4.3, the previously detected point source 2FGL J0526.6+4308 is removed from the model. This source is found at the position RA (J2000)= 05$^h$26$^m$39.7$^s$, Dec (J2000)= 43$^\circ$08$'$48.8$''$ within the remnant's shell. Therefore, no source at the positions of the SNRs is originally included in this part of the analysis. The resulting model is referred to as the null hypothesis and it is thus a model of the background.

The significance of a source is estimated by means of a test statistic (TS) defined as $-2$ log($L_0/L$), where $L_0$ and $L$ are the maximum likelihood values for the null hypothesis and for a model including the additional source, respectively \citep{mat96}. In the limit of a large number of counts, despite some caveats \citep{protassov2002}, the detection significance of the source is roughly given by $\sqrt{\mbox{TS}}$, and therefore TS $=25$ is usually considered the threshold value for detection (corresponding to a significance of $5\sigma$ for one degree of freedom).

The best-fit models that result from the maximization procedures are used to generate maps of the background emission which are smoothed with a boxcar of length $0^{\circ}.5$. When subtracted from the corresponding smoothed count maps, the resulting residuals maps can be used to visualize the disagreement between the observed counts and the model. These background-subtracted maps are obtained for different energy bands and then divided by the square root of the background, resulting in the signal-to-noise maps that are shown in Figs. \ref{fig1} and \ref{fig2} for the two sources.

Representative radio contours of G296.5+10.0 and G166.0+4.3 obtained from observations of the Green Bank catalog of radio sources \cite[GB6, 4860 MHz,][]{gregory1996} and the Westerbork Northern Sky Survey \cite[WENSS, 325 MHz,][]{rengelink1997}, respectively, are overlaid onto the signal-to-noise maps. Residual emission coincident with the positions of the sources is apparent. Maximum likelihood analyses are next performed in order to quantify the significance and test the spatial morphology and spectral parameters of the emission.

\subsubsection{Morphology}\label{morphology}
As shown in Figs. \ref{fig1} and \ref{fig2} only events above 0.5 GeV are used for morphology studies, to take advantage of the narrower PSF at higher energies. Several hypotheses for the morphology of the residual gamma-ray emission are tested.

\subsubsubsection{G296.5+10.0}
Fig. \ref{fig1} shows the signal-to-noise maps obtained for the region around this SNR with the corresponding radio contours of the bipolar shell for two energy bands: 2-7 GeV and 7-100 GeV. In the energy range above 2 GeV the emission seems to come in part from a region towards the interior of the remnant, and it becomes more prominent towards the western limb of the SNR at the highest energies. This excess seems to follow the western radio emission.

In order to evaluate the significance of the emission, a `TS map' is calculated above 500 MeV for the region near the SNR. The map is obtained by evaluating the TS value of a point source that is moved in a predefined grid. Based on these results, the gamma-ray morphology is assessed by fitting different spatial models for G296.5+10.0: (a) a spatial template obtained from the radio GB6 observation, (b) a uniform disk template and (c) a point-source located at the position of maximum value in the TS map. A simple power-law model is assumed for the energy spectrum. For the disk, the TS values with respect to the null hypothesis are evaluated for different locations and sizes. The radius of the disk template is systematically increased in steps of 0$^{\circ}$.1 and the position of its centroid is changed within the interior of the remnant shell.

The TS values obtained for the radio template, uniform disk and point-source hypotheses are 28, 36 and 25, respectively, as shown in Table \ref{table1}. The best-fit disk radius and centroid location are 0$^{\circ}$.6$\pm 0^{\circ}.1$ and RA (J2000)= 12$^h$09$^m$, Dec (J2000)= -52$^\circ$27$'$. The error of the centroid is $0^{\circ}.2$ at 95$\%$ confidence level. Although the fit with the disk template has a higher TS with respect to the fit with the radio template (the difference being $\Delta$TS $= 8$), the corresponding model has 3 additional parameters and therefore both can be considered statistically similar descriptions of the source morphology. The disk hypothesis is an improvement with respect to the point-source ($\Delta$TS $=11$, see Table \ref{table1}), the analysis thus indicates that the source extension is detected at the $3\sigma$ level \citep{lande2012}. The disk hypothesis will then be adopted for spectral analysis here and as explained later on the systematic uncertainties of changing the hypothesis for the source morphology will be considered.

As an additional test to evaluate the amount of background emission that may have not been accounted for in the model, new fits are performed by moving the disk template in different azimuthal positions outside and around the radio contours. No significant emission is detected outside/around the shell of the SNR.

\subsubsubsection{G166.0+4.3}
Fig. \ref{fig2} shows a correspondence between the gamma-ray signal and the radio contours of the shell of the SNR for two energy bands: 0.5-2 GeV and 2-100 GeV. For the latter energy band, the gamma-ray emission seems to become more significant near the west, where the extended wing is located.

Next, two hypotheses for the gamma-ray emission are tested: (a) one following a spatial template obtained from the WENSS 325MHz observation, and (b) a point-source hypothesis, located at the position of the 2-year LAT Catalog source 2FGL J0526.6+4308. A simple power-law is used to model the energy spectrum in both cases. Table \ref{table1} shows the TS values obtained for the two spatial hypotheses with respect to the null hypothesis (no emission from G166.0+4.3).

The TS value for the radio contour hypothesis\footnote{A similar TS value is obtained for a disk morphology hypothesis for G166.0+4.3. Since this hypothesis introduces 3 additional free parameters, it is not used here.} (91) increases with respect to the TS value for the point source (83). The radio template scenario for the emission may be considered a slight improvement in the fit with respect to the point source and has been adopted for the spectral analysis.

\subsubsubsection{Association}
There are no known radio pulsars near the position of the SNRs. G296.5+10.0 is known to harbor a radio-quiet neutron star left from the supernova explosion \citep{zavlin2000}, but these compact objects do not show gamma-ray emission. Moreover, the results shown here support the presence of an extended source of gamma-ray emission. The CRATES catalog of flat spectrum radio sources \citep{healey2007} contains a few sources near the contours of G296.5+10.0. The three closest are located at the positions $(l,b) \sim (295^{\circ}.66,+10^{\circ}.12)$, $(296^{\circ}.45,+10^{\circ}.50)$, $(296^{\circ}.34,+10^{\circ}.59)$. These sources are found towards the north of the shell, on average around $0^{\circ}.7$ from the peak of the 7-100 GeV signal-to-noise map. In what follows, then, the gamma-ray emission seen in the direction of G166.0+4.3 and G296.5+10.0 is assumed to be associated with the SNRs.

\subsubsection{Spectral Analysis}
A binned likelihood analysis is performed in the energy band 0.2-100 GeV for both sources using the best-fit spatial templates found in Section \ref{morphology}. Using 10 energy bins per decade in the exposure map calculation and assuming power-law spectra, the resulting TS values for the sources G166.0+4.3 and G296.5+10.0 in this energy band are 136 and 36, corresponding to significances of roughly $11\sigma$ and $5\sigma$ over the background for 2 and 5 degrees of freedom, respectively.

In order to probe for curvature in the spectra, a log-parabolic spectral shape, $dN/dE = N_0 (E/E_b)^{-(\alpha + \beta \mbox{\small log} (E/E_b)}$ is tested in the fit for both sources, but no significance of a deviation from a power-law spectral distribution can be claimed for either one. The best-fit spectral parameters are summarized in Table \ref{table2}.

The spectral indices are very different for both sources, G166.0+4.3 having a much steeper spectrum. No significant emission for this source is detected above 10 GeV. On the other hand, G296.5+10.0 would be among the dimmest gamma-ray emitting SNRs detected so far, with a luminosity of $(3.3 \pm 0.9)\times10^{33}$ ergs/s (1-100 GeV), about three times that of the Cygnus Loop \citep{katagiri2011}.

Gamma-ray spectral energy distributions (SEDs) for G166.0+4.3 and G296.5+10.0 are obtained with maximum likelihood fits for 3 logarithmically-spaced energy intervals in the bands 0.2-8.3 GeV and 0.2-100 GeV, respectively. If the significance of a detection within a bin is less than $3\sigma$ an upper limit for the flux is derived. Several sources of systematic errors are considered for these bins: a) the effect of changing the morphology of the emission as explained in Section \ref{morphology}; b) the uncertainty of the Galactic diffuse emission, which is evaluated by artificially varying the best-fit value of the normalization of the Galactic level by $\pm 6\%$ in each bin (as done by Abdo et al. 2009a); and c) the systematic uncertainty in the effective area, which is energy-dependent and given by $10\%$ at 100 MeV, $5\%$ at 560 MeV and $20\%$ at 10 GeV \citep{abdo2009b}. Fig. \ref{fig4} shows the broadband SEDs and gamma-ray data points with the resulting statistical and systematic errors added in quadrature.

\section{Emission Model}\label{model}
The electron distributions used here are broken power-laws of the form $$N_{e}(\epsilon)= K \epsilon^{-s} \left(1+\left(\frac{\epsilon}{\epsilon_{br}}\right)^2\right)^{-\delta}$$ extending up to a maximum energy $\epsilon_{\mbox{\tiny max}}$. The break ($\epsilon_{br}$) and maximum energy as well as the parameter $\delta$ are varied to fit the shape of the gamma-ray spectra. The proton distributions used to model the data are simple power-laws, $N_p \propto \epsilon ^{-s_p}$, based on the fact that the gamma-ray spectra of the sources are also power-laws.

Instead of applying a model to follow the evolution of the SNRs and predict the break energies and magnetic field values, these parameters are derived under the assumption that the nature of the gamma-ray emission is either leptonic or hadronic, in other words, the necessary conditions that produce either outcome are explored. It is noted also that a full exploration of the parameters that reproduce the gamma-ray data is beyond the scope of this work and only a representative set of parameters is presented for each source.

The volume occupied by the particles is the volume of the SNR and the supernova explosion energy is set to $E_{SNR}=10^{51}$ ergs. Gamma-ray emission mechanisms include IC of cosmic microwave background (CMB) photons, bremsstrahlung emission from electrons and neutral pion decay, which depend on the target material density, denoted as $n_H$ (cm$^{-3}$). For both SNRs, leptonic-dominated and hadronic-dominated scenarios are considered. The gamma-ray flux resulting from hadron interactions is calculated as in Kamae et al. \citep{kamae2006}

\subsection{G296.5+10.0}
As noted before, the spectral index of electrons is fixed from radio observations to $s=2$. The remnant is approximated as a sphere of radius 0$^{\circ}$.7 (at a source distance of 2.1 kpc this is equivalent to 26 pc).

\subsubsection{Leptonic-dominated scenario}
The gamma-ray emission can be accounted for by IC on the CMB, as shown in Fig. \ref{fig4}, with a `softening' given by $\delta=0.45$, an electron break energy $\epsilon_{br}\simeq 100$ GeV, a magnetic field of 60 $\mu$G, a total electron energy of $5.5\times10^{48}$ ergs and ambient density 0.05 cm$^{-3}$ or less. Although not shown, a bremsstrahlung-dominated scenario results with a magnetic field $B \simeq 100 \,\mu$G, total electron energy $2.4\times10^{48}$ ergs and ambient density 1 cm$^{-3}$. The maximum electron energy adopted is $\epsilon_{\mbox{\tiny max}}= 1.5$ TeV.

\subsubsection{Hadronic-dominated scenario}
Given the relatively low statistics of the observation and the high systematic errors at several hundreds of MeV, the gamma-ray index is not well constrained. If a standard power-law proton distribution is assumed with index $s_p=2$, a total cosmic-ray proton energy of $1.1\times10^{49}(n_H/1\,\mbox{cm}^{-3})^{-1}$ ergs is needed to account for the high-energy SED. The proton energy spectrum shown in Fig. \ref{fig5} extends from the pion production threshold to 250 GeV, although \emph{Fermi} LAT data are not inconsistent with higher proton energies, as there is no evidence for a spectral cutoff. The magnetic field is $240\,\mu$G in this case and the total electron energy is $6.6\times 10^{47}$ ergs, decreasing with increasing magnetic field.

\subsection{G166.0+4.3}
The spectral index of electrons, again from radio observations, is fixed at $s=1.7$. The larger estimated distance and smaller angular extension of G166.0+4.3 compared to G296.5+10.0 yield, however, a similar intrinsic remnant size. The remnant is also approximated as a sphere of radius 26 pc.

\subsubsection{Leptonic-dominated scenario}
In the case that the emission is mainly from IC on the CMB, an electron spectral `softening' given by the parameter $\delta=0.5$, particle break energy and maximum energy of 50 GeV and 500 GeV, respectively, are necessary. The magnetic field and total electron energy are $\simeq 5 \,\mu$G and $8.7\times10^{49}$ ergs, respectively. The value for the density used for calculating the bremsstrahlung emission is $n_H=0.1$ cm$^{-3}$, although a value lower than this is probably more realistic, as X-ray observations show \citep{burrows1994}.

\subsubsection{Hadronic-dominated scenario}
The gamma-ray data requires a steep cosmic-ray proton spectrum (the photon spectral index is 2.7). Fig. \ref{fig5} shows a scenario where the gamma-ray emission is accounted for by a power-law proton distribution with index $s_p=2.7$. The magnetic field is 50 $\mu$G and the total proton energy $(2.8_{-1.2} ^{+2.2})\times10^{50}(n_H/1\,\mbox{cm}^{-3})^{-1}$ ergs, considering the uncertainty in the distance \cite[$4.5\pm{1.5}$ kpc,][]{giacani2000}. The maximum particle energy is not well-constrained since the contribution from the highest energy particles is less important for such a steep distribution, the value used in Fig. \ref{fig5} is $\simeq 500$ GeV. Fig. \ref{fig5} shows a scenario with $n_H =1$ cm$^{-3}$, a total proton energy of $\simeq 30\%$ of the available SNR energy, a total electron energy of $3.9\times 10^{48}$ ergs and a bremsstrahlung flux calculated also with an ambient density $n_H =1$ cm$^{-3}$. Note that if instead a much lower ambient density is adopted as implied by X-ray observations \cite[$n_H \sim 0.01$ cm$^{-3}$,][]{burrows1994}, the gamma-ray observations require an unrealistically large total proton energy.

\section{Discussion and Conclusions}\label{discussion}
Excess gamma-ray emission in the region of the SNRs G166.0+4.3 and G296.5+10.0 is revealed by analysis of accumulated observations from the \emph{Fermi} LAT. In the case of G166.0+4.3, a hypothesis for the emission from the radio shell and wing (325 MHz) is slightly preferred over the previously detected point source located in the shell. For SNR G296.5+10.0 extended gamma-ray morphology is also preferred over point source emission (see Table \ref{table1}).

The gamma-ray spectra of the sources are very different. They can be described by simple power-laws with no clear evidence of curvature. The spectrum of G166.0+4.3 is very steep (photon index $2.7 \pm{0.1}$, statistical errors only) while that of G296.5+10.0 (photon index $1.85 \pm{0.13}$, statistical errors only) is somewhat harder than that expected for a standard test-particle DSA spectrum with a power-law index of 2. The broad-band SEDs of both sources can be either of leptonic or hadronic origin.

The hadronic interpretation for the emission from G166.0+4.3 presents, however, some difficulties. The steep proton spectrum required to account for the gamma-ray SED is hard to understand in the context of standard shock acceleration, and the ambient density required is much higher than that obtained from X-ray observations \citep{burrows1994}. The contradiction could be solved if the SNR interacts with irregular ISM with clumps of high-density material located in a low-density environment, as has been proposed for other SNRs \citep{castro2010}. It is believed, based on HI observations of the region, that the SNR has interacted with material in its environment \citep{landecker1989}. However, it is unclear whether the density of the target material is enough to account for the flux and even if it was, it would still be hard to explain the steep spectrum.

Another possibility is that the emission from G166.0+4.3 is leptonic. As shown in Fig. \ref{fig4}, the IC on the CMB level is enough to account for the \emph{Fermi} LAT points with reasonable SNR parameters, and it is compatible with the low ambient density implied by X-ray observations, which affects the bremsstrahlung level only. In this scenario, the observed steep gamma-ray spectrum is produced mainly by the highest-energy synchrotron-emitting electrons in the tail of the distribution, above a particle break of $\simeq 50$ GeV according to the model. The leptonic emission scenario is compatible with the observed correspondence between the radio contours of the source and the gamma-ray maps, as seen in Fig. \ref{fig2}.

From these considerations, the leptonic emission scenario for the SED of G166.0+4.3 is found to be a more natural explanation. Future observations might allow performing a more detailed analysis considering different electron populations, perhaps disentangling the properties of particles in the wing and shell, as well as studying spatial variations of spectral parameters.

\begin{figure}[ht]
\centering
 \subfigure[ ]{\includegraphics[width=8cm,height=6cm]{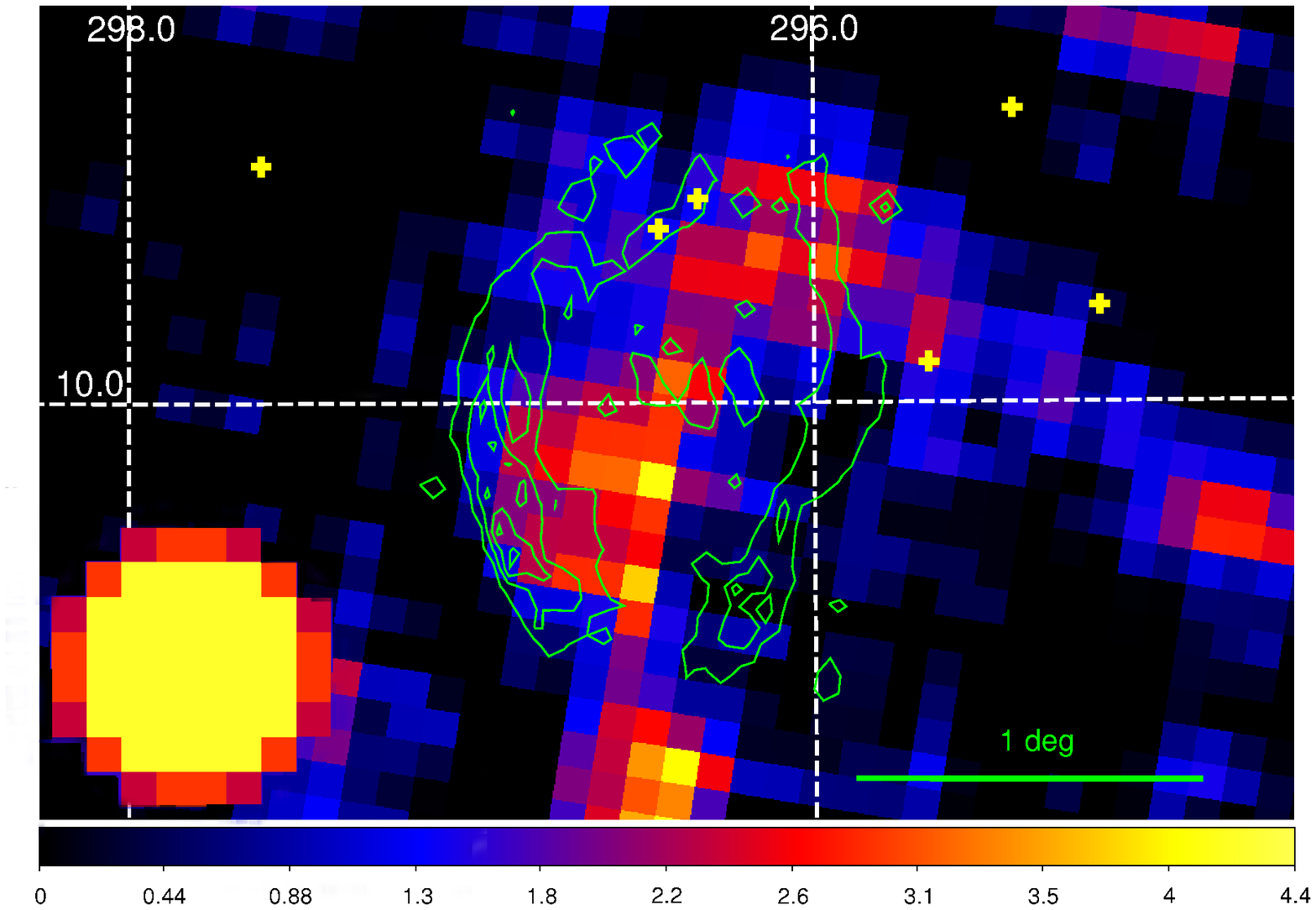}}
 \subfigure[ ]{\includegraphics[width=8cm,height=6cm]{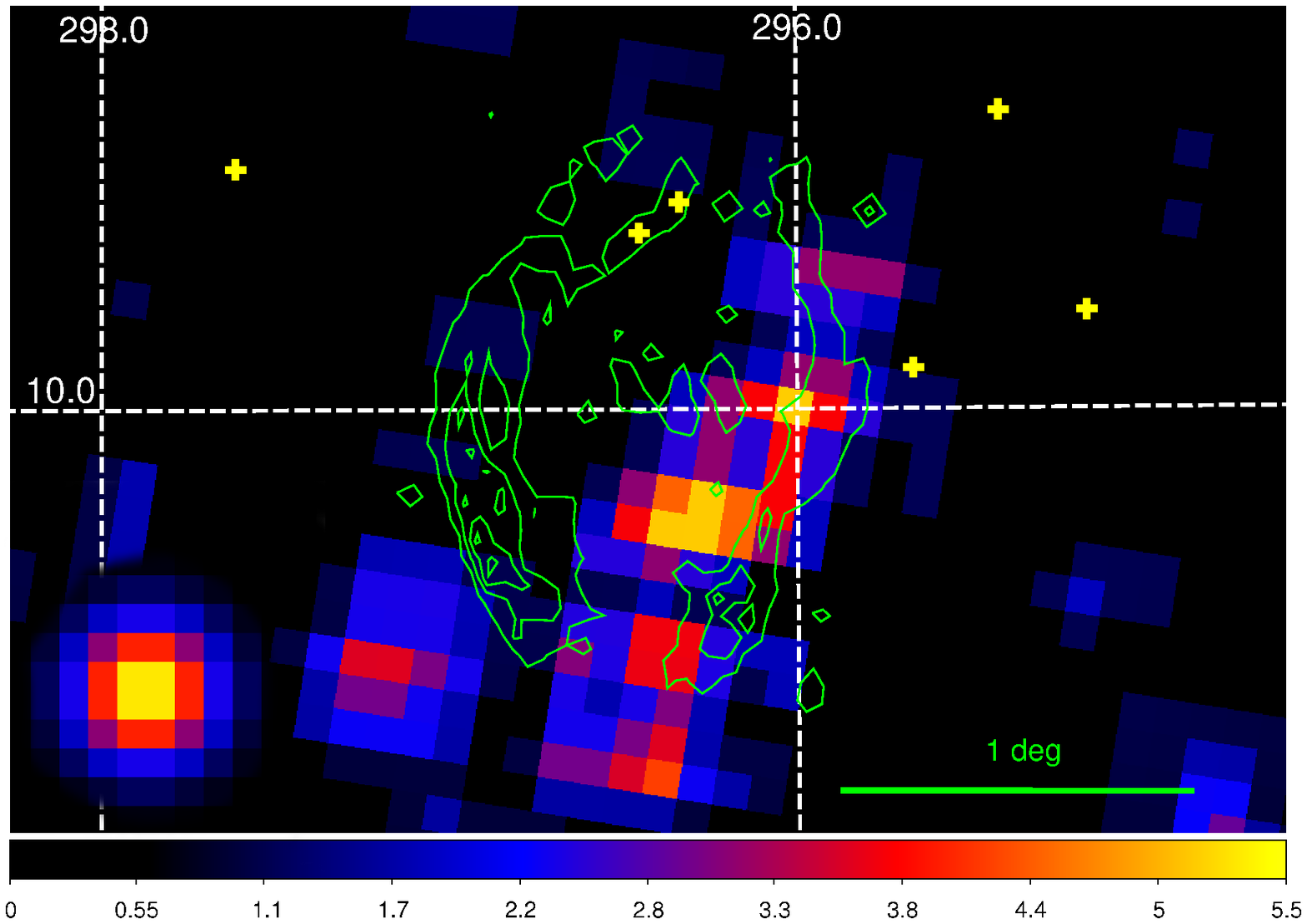}}
\caption{Signal-to-noise maps in the region of the SNR G296.5+10.0 smoothed with a boxcar of length 0$^\circ$.5 in two energy bands: (a) 2-7 GeV, and (b) 7-100 GeV (see text), and green contours of a GB6 4860 MHz observation of the SNR. The position of the CRATES sources are represented by (yellow) crosses and Galactic coordinates are shown in degrees. The insets in the bottom left corner show the smoothed PSF.\label{fig1}}
\end{figure}

\begin{figure}[ht]
\centering
 \subfigure[ ]{\includegraphics[width=8cm,height=6cm]{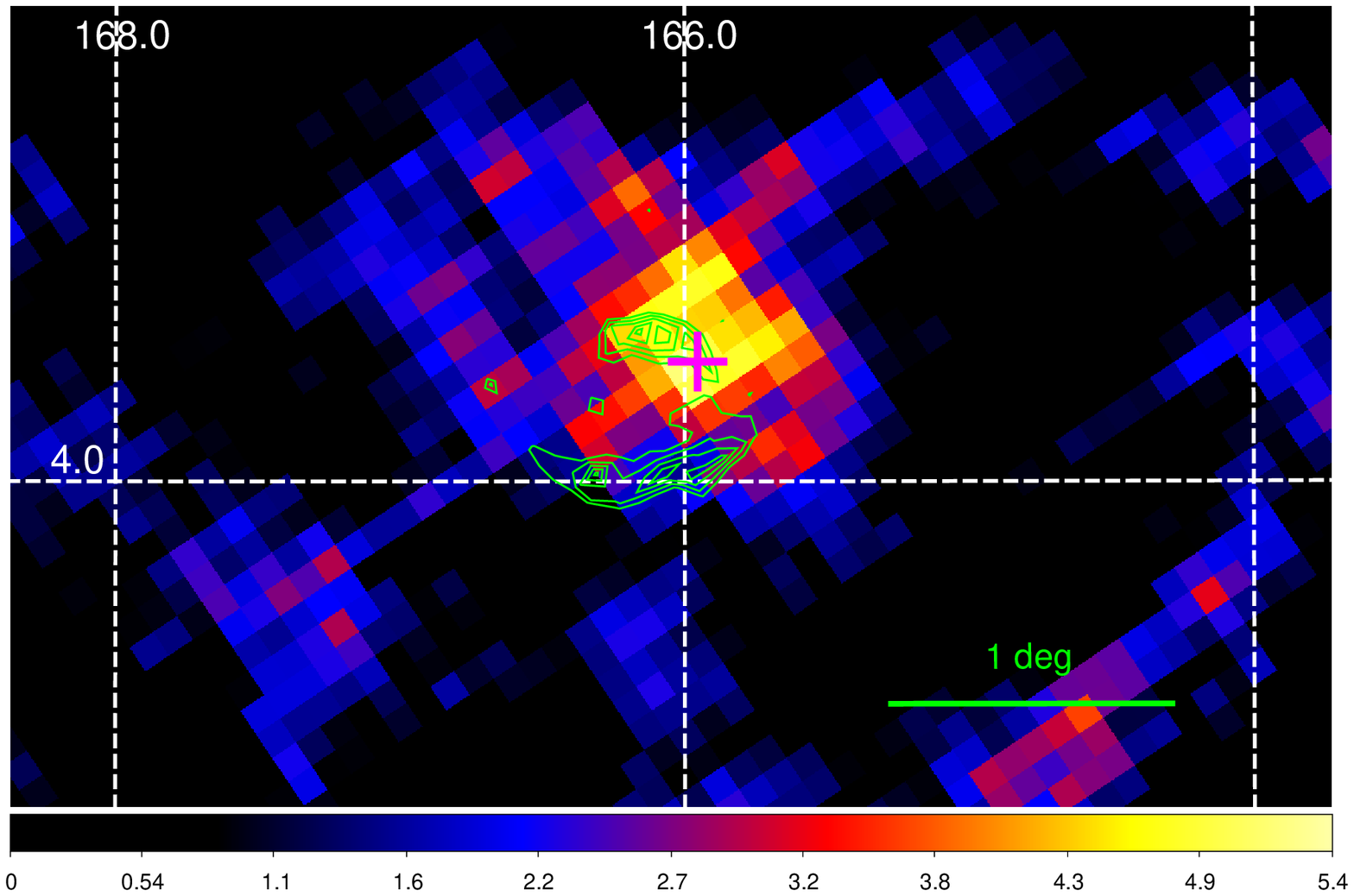}}
 \subfigure[ ]{\includegraphics[width=8cm,height=6cm]{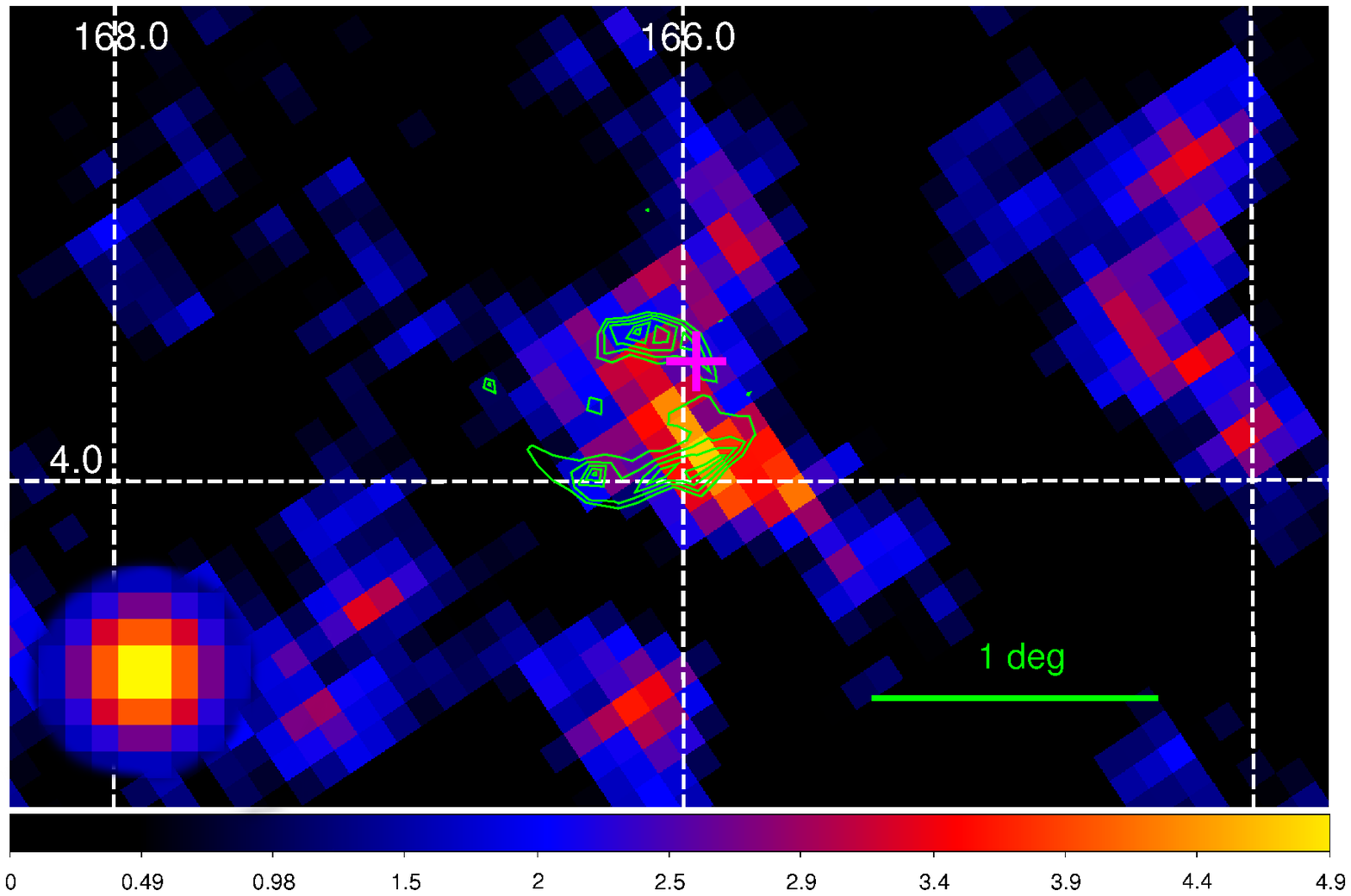}}
\caption{Signal-to-noise maps in the region of the SNR G166.0+4.3 smoothed with a boxcar of length 0$^\circ$.5 in two energy bands: (a) 0.5-2 GeV and (b) 2-100 GeV. Overlaid are green contours of the radio WENSS observation (325 MHz) of the SNR. The magenta cross indicates the position of the source 2FGL J0526.6+4308. Galactic coordinates are shown in degrees and the PSF is only shown for the higher energy map. \label{fig2}}
\end{figure}

\begin{figure}[h]
\begin{center}
\subfigure[ ]{\includegraphics[width=8cm,height=5cm]{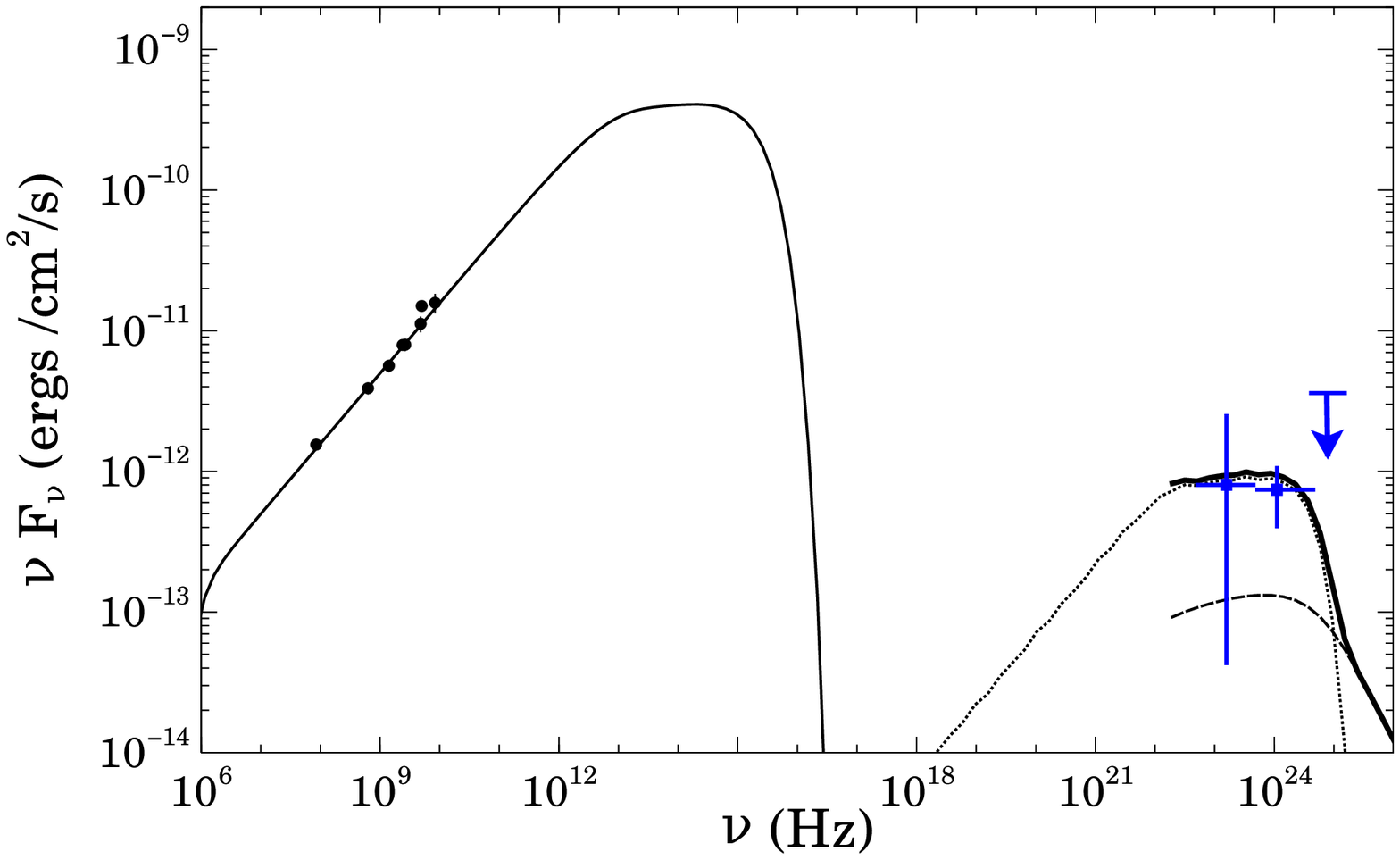}}
\subfigure[ ]{\includegraphics[width=8cm,height=5cm]{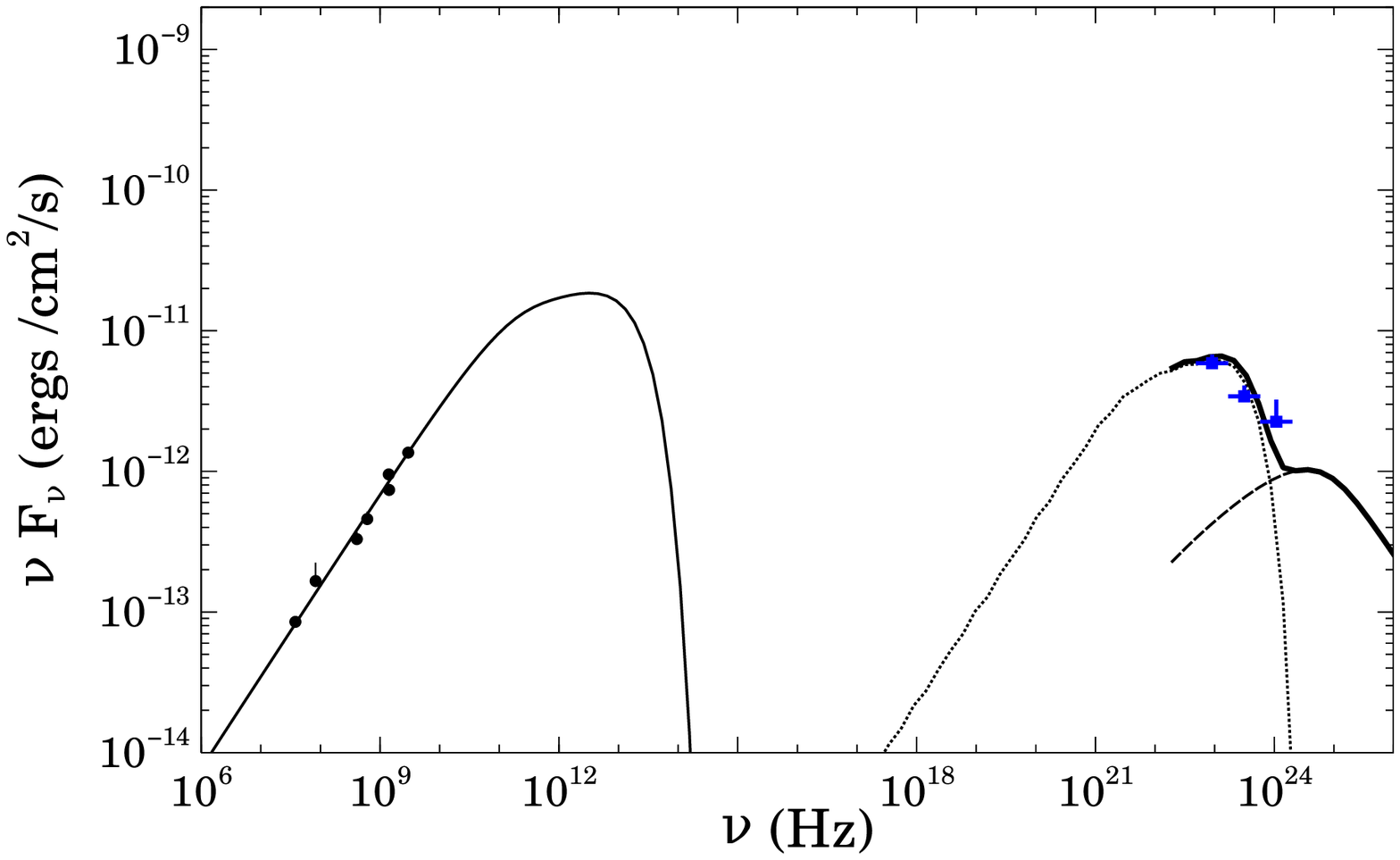}}
\caption{Leptonic emission model: (a) G296.5+10.0, (b) G166.0+4.3. The emission components are: synchrotron (solid line), IC-CMB (dotted line) and non-thermal bremsstrahlung (dashed line, shown only above $\sim 2\times10^{22}$ Hz). The dark solid line represents the total gamma-ray emission. Blue squares are obtained from the LAT observation presented in this paper.}
\label{fig4}
\end{center}
\end{figure}

\begin{figure}[h]
\begin{center}
\subfigure[ ]{\includegraphics[width=8cm,height=5cm]{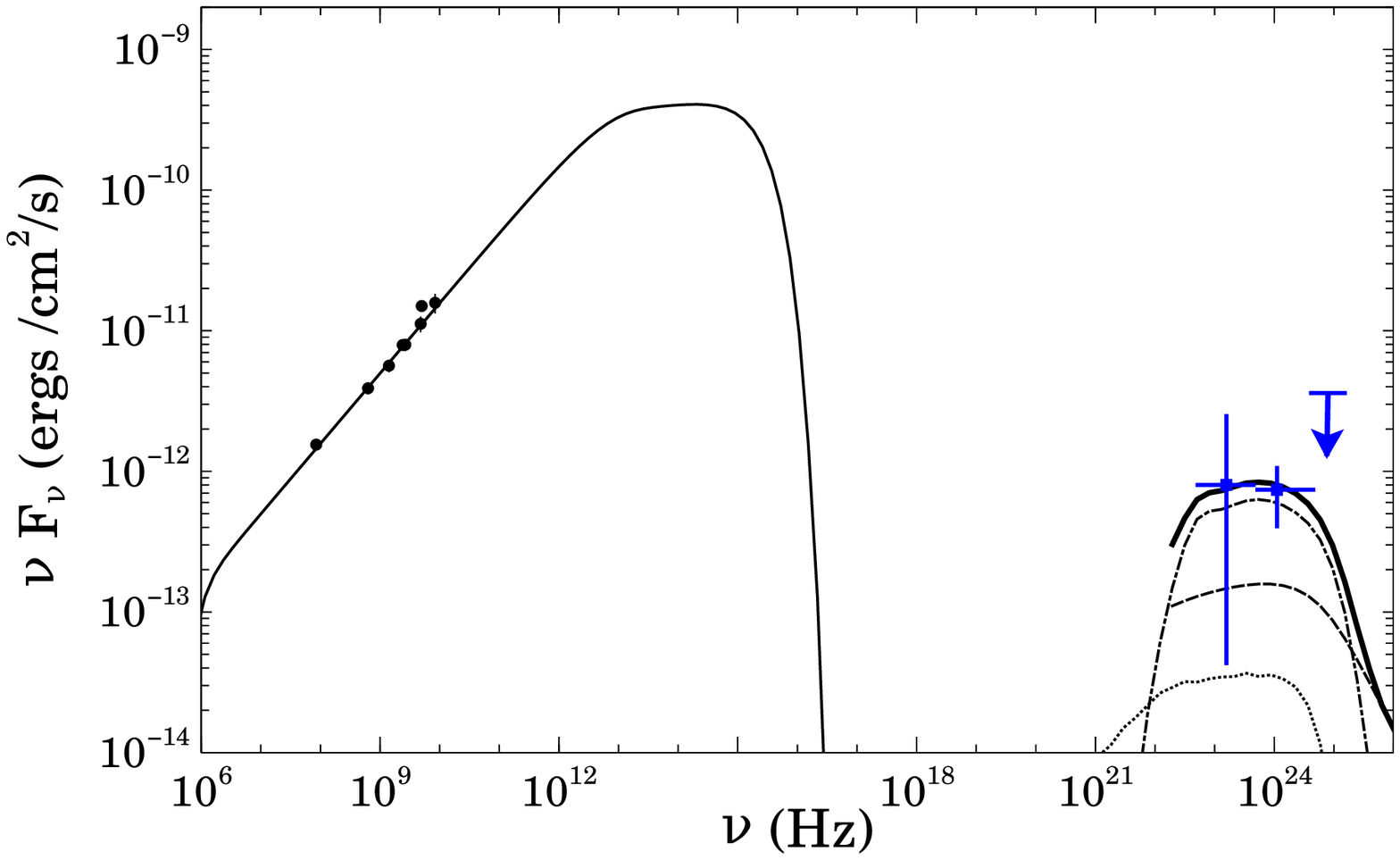}}
\subfigure[ ]{\includegraphics[width=8cm,height=5cm]{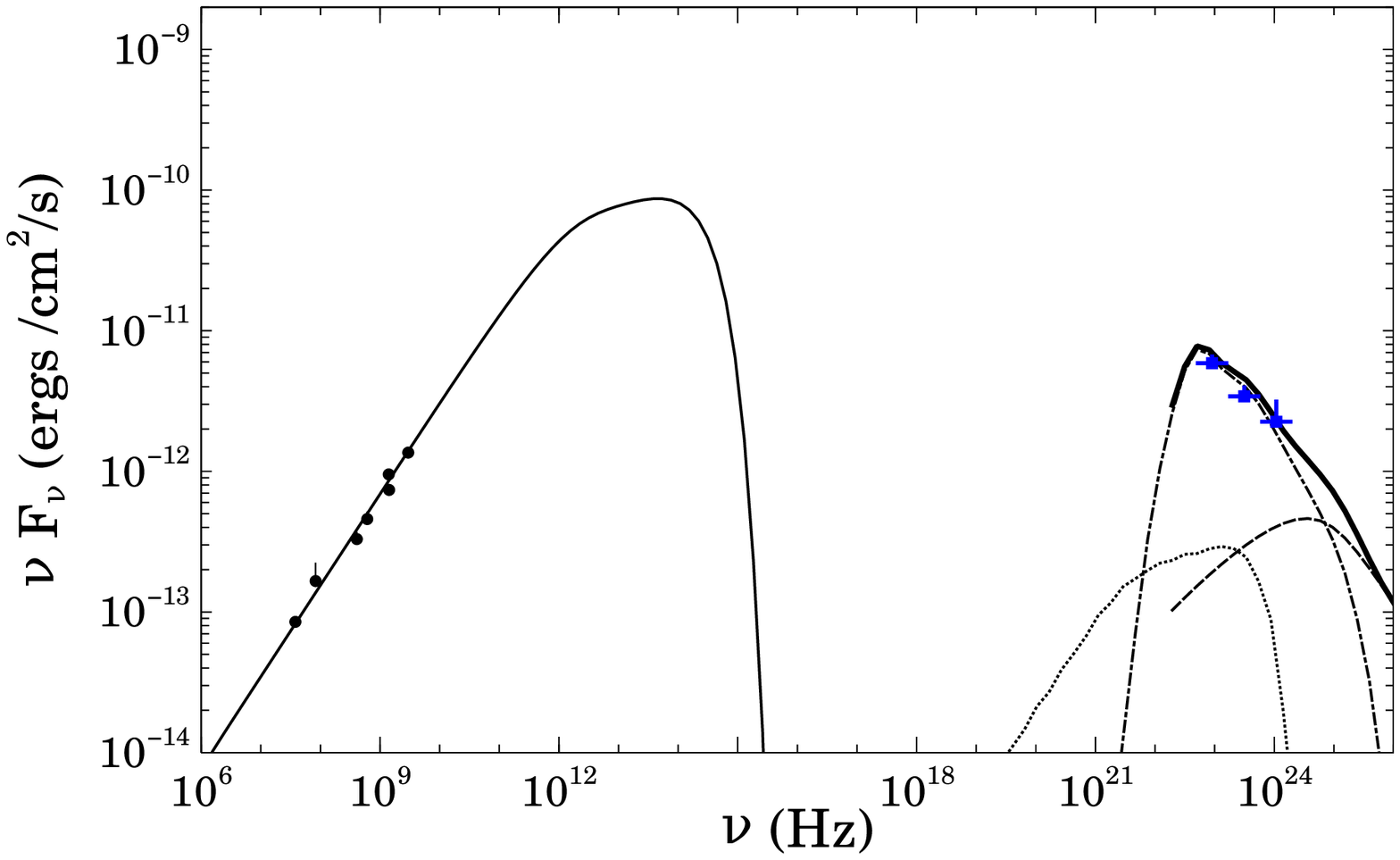}}
\caption{Hadronic emission model: (a) G296.5+10.0, (b) G166.0+4.3. The emission components are: synchrotron (solid line), IC-CMB (dotted line), non-thermal bremsstrahlung (dashed line, shown only above $\sim 2\times10^{22}$ Hz) and hadronic emission from $\pi^0$ decay, from proton-proton interactions (dash-dotted line). The dark solid line represents the total gamma-ray emission. Blue squares are obtained from the LAT observation presented in this paper.}
\label{fig5}
\end{center}
\end{figure}

In the case of G296.0+10.0, the hypothesis of a uniform disk, containing most of the radio shell, of radius 0$^{\circ}$.6$\pm 0^{\circ}.1$ and centroid location RA (J2000)= 12$^h$09$^m$, Dec (J2000)= -52$^\circ$27$'$ (Galactic coordinates (l,b) $\sim (296^{\circ}.4,+9^{\circ}.88)$) was adopted here. However, the residuals in Fig. \ref{fig1} suggest that the gamma-ray morphology might be more complicated than the scenarios explored here would imply.

At the highest energies, the peak of the signal-to-noise maps (see Fig. \ref{fig1}) is located to the east of the western radio contours, where synchrotron emission is also present \citep{gregory1996}. This is interesting since both the X-ray and radio fluxes are highest in the eastern hemisphere. It is possible, in the context of a hadronic scenario, that at least part of the gamma-ray emission is produced by high-energy particles that interact with the dense ($n_H=13$ cm$^{-3}$) southwestern cloud seen \citep{giacani2000} in contact with the shell, although there is another dense cloud also interacting with the shell in the east. The SED parameters derived for leptonic and hadronic scenarios are both physically reasonable.

The required mean density for a bremsstrahlung-dominated scenario for G296.5+10.0 ($n_H=1$ cm$^{-3}$) might be in conflict with previous observations \cite[$0.24$ cm$^{-3}$ and $0.08$ cm$^{-3}$;][respectively]{kellet1987,matsui1988}. Roger et al. \citep{roger1988} argue that this SNR is in the adiabatic (Sedov-Taylor) phase of evolution which is consistent with the remnant size, the observed low ambient density and thus an IC origin for the gamma-ray emission. Furthermore, the particle break and magnetic field required by the gamma-ray data ($\epsilon_{br}\simeq 100$ GeV and $B\simeq 60\, \mu$G) are consistent with a synchrotron cooling break \cite[see, e.g.,][]{tanaka2008} and a reasonable SNR age \cite[$\sim 10^4$ yr,][]{vasisht1997}. If the emission from this SNR is mainly leptonic, an IC origin seems then more plausible than a bremsstrahlung origin.

\acknowledgments
This research has made use of NASA's Astrophysical Data System and of the SIMBAD database, data from the \emph{WENSS} team \citep{rengelink1997} and from the NRAO. Financial support from Universidad de Costa Rica is acknowledged. Valuable observations from the anonymous referee helped improve the quality of this work substantially.

\begin{table}
\begin{center}
\caption{Test Statistic for Different Spatial Models Compared to the Null Hypothesis (0.5-100 GeV)\label{table1}}
\begin{tabular}{|c|c|c|}
\tableline\tableline
 & Test Statistic & Additional Degrees of Freedom\\
\tableline
G296.5+10.0 & &\\
Null hypothesis (background only) & 0 & 0\\
Uniform disk$^{a}$ & 36 & 5\\
Point source & 25 & 4\\
4860 MHz radio continuum & 28 & 2\\
\tableline
G166.0+4.3 & &\\
Null hypothesis (background only) & 0 & 0\\
Point source$^{b}$ & 83 & 2\\
325 MHz radio continuum & 91 & 2\\
\tableline
\end{tabular}
\tablenotetext{a} {Best fit radius and centroid position 0$^{\circ}$.6$\pm 0^{\circ}.1$, RA (J2000)= 12$^h$09$^m$, Dec (J2000)= -52$^\circ$27$'$ ($\pm 0^{\circ}$.2 at 95$\%$ confidence level).}
\tablenotetext{b} {Found in the 2-year LAT catalog \citep{nolan2012}.}
\end{center}
\end{table}

\begin{table}
\begin{center}
\caption{Best-Fit Spectral Parameters$^{a}$ (0.2-100 GeV)\label{table2}}
\begin{tabular}{|c|c|c|c|}
\tableline\tableline
& Spectral Index & Integrated Photon Flux (cm$^{-2}$s$^{-1}$) & TS\\
\tableline
G296.5+10.0 & & &\\
 & $1.85\pm{0.13}$ & $(3.1\pm{0.9})\times 10^{-9}$ & 36\\
\tableline
G166.0+4.3 & & &\\
 & $2.7\pm{0.1}$ & $(1.84\pm{0.18})\times 10^{-8}$ & 136\\
\tableline
\end{tabular}
\tablenotetext{a} {Assuming power-law spectra and best-fit spatial template. Only statistical errors are shown.}
\end{center}
\end{table}

\clearpage

\end{document}